\documentclass[prb,twocolumn,superscriptaddress,nobibnotes]{revtex4}

\usepackage[dvips]{graphicx}
\usepackage{amsmath}
\usepackage{pslatex}
\usepackage{amssymb}
\usepackage{amsfonts}
\usepackage{color}

\usepackage{hhline}
\usepackage{threeparttable}
\usepackage{supertabular}
\usepackage{multirow}
\usepackage{tabularx}

\newcolumntype{Y}{>{\centering\arraybackslash}X}

\begin{document}

\pagenumbering{arabic}

\title{Linear and nonlinear optical spectroscopy of a strongly-coupled microdisk-quantum dot system}

\author{Kartik Srinivasan}
\email{kartik@caltech.edu}
\affiliation{Center for the Physics
of Information, California Institute of Technology, Pasadena, CA}
\author{Oskar Painter}
\email{opainter@caltech.edu}
\affiliation{Thomas J. Watson, Sr., Laboratory of Applied Physics, California Institute of Technology, Pasadena, CA 91125}

\date{\today}

\begin{abstract}
  A fiber taper waveguide is used to perform direct optical spectroscopy of a microdisk-quantum-dot system, exciting the
  system through the $\textit{photonic}$ (light) channel rather than the $\textit{excitonic}$ (matter) channel.  Strong
  coupling, the regime of coherent quantum interactions, is demonstrated through observation of vacuum Rabi splitting in
  the transmitted and reflected signals from the cavity. The fiber coupling method also allows the examination of the
  system's steady-state nonlinear properties, where saturation of the cavity-QD response is observed for less than one
  intracavity photon.
\end{abstract}

\maketitle

Cavity quantum electrodynamics\cite{ref:Kimble2}, the study of coherent quantum interactions between the
  electromagnetic field and matter inside a resonator, has received attention as both a testbed for ideas in quantum
  mechanics and also as a building block for applications in the field of quantum information
  processing\cite{ref:Mabuchi}. The canonical experimental system studied in the optical domain is a single alkali atom
  coupled to a high-finesse Fabry-Perot cavity. The tremendous progress made in this
  system\cite{ref:Kimble2,ref:Mabuchi,ref:Hood2,ref:Hennrich,ref:Boca} has been complemented by recent
  research involving trapped ions\cite{ref:Keller}, chip-based microtoroid cavities\cite{ref:Aoki1},
  integrated microcavity-atom-chips\cite{ref:reichel1}, nanocrystalline quantum dots coupled to microsphere cavities
  \cite{ref:Park_Y}, and semiconductor quantum dots embedded in micropillars, photonic crystals, and
  microdisks\cite{ref:Reithmaier,ref:Yoshie3,ref:Peter}. The latter system has been of particular interest due to its
  potential simplicity and scalability. In contrast to preceding work with semiconductor systems, which has focused on
  photoluminescence measurements \cite{ref:Reithmaier,ref:Yoshie3,ref:Peter,ref:Hennessy3,ref:Press}, here we use a
  fiber taper waveguide to perform direct optical spectroscopy of a microdisk-quantum-dot system, exciting the system
  through the $\textit{photonic}$ (light) channel rather than the $\textit{excitonic}$ (matter) channel.  Strong
  coupling, the regime of coherent quantum interactions, is demonstrated through observation of vacuum Rabi splitting in
  the transmitted and reflected signals from the cavity. The fiber coupling method also allows us to examine the
  system's steady-state nonlinear properties, where we see a saturation of the cavity-QD response for less than one
  intracavity photon. The excitation of the cavity-QD system through a fiber optic waveguide is key for applications
  such as high-efficiency single photon sources\cite{ref:Michler,ref:Santori2}, and to more fundamental studies of the
  quantum character of the system\cite{ref:Birnbaum}.

In the most simplified picture, cavity quantum electrodynamics (cQED) consists of a single two-level atom (or
equivalent) coupled to an electromagnetic mode of a cavity.  A more realistic picture includes dissipative processes,
such as cavity loss and atomic decoherence, and excitation of the system, either through the atomic or photonic channel.
The observed system response is dependent on both which channel is excited, and what signal is measured.  Previous
demonstrations of strong coupling between semiconductor microcavities and quantum dots
(QDs)\cite{ref:Reithmaier,ref:Yoshie3,ref:Peter,ref:Hennessy3,ref:Press} used non-resonant optical pumping to excite
the QD stochastically and photoluminescence (PL) to probe the system behavior.  In this work we excite the system
coherently through the photonic channel, and detect signatures of cavity-QD coupling in the resonant optical response.
Such optical spectroscopy is commonplace in atom-Fabry-Perot systems\cite{ref:Kimble2}, but is more problematic in
semiconductor microcavities due to the comparative difficulty in effectively coupling light into and out of sub-micron
structures.  To effectively interface with the cavity, we use an optical fiber taper waveguide\cite{ref:Knight}.  Fiber
tapers are standard glass optical fibers that have been heated and stretched to a diameter at or below the
wavelength of light, at which point the evanescent field of the guided mode extends into the surrounding air and allows
the taper to function as a near-field optic\cite{ref:Spillane2,ref:Srinivasan7,ref:Srinivasan9,ref:Aoki1}.

\begin{figure}[t]
\begin{center}
\includegraphics[width=\linewidth]{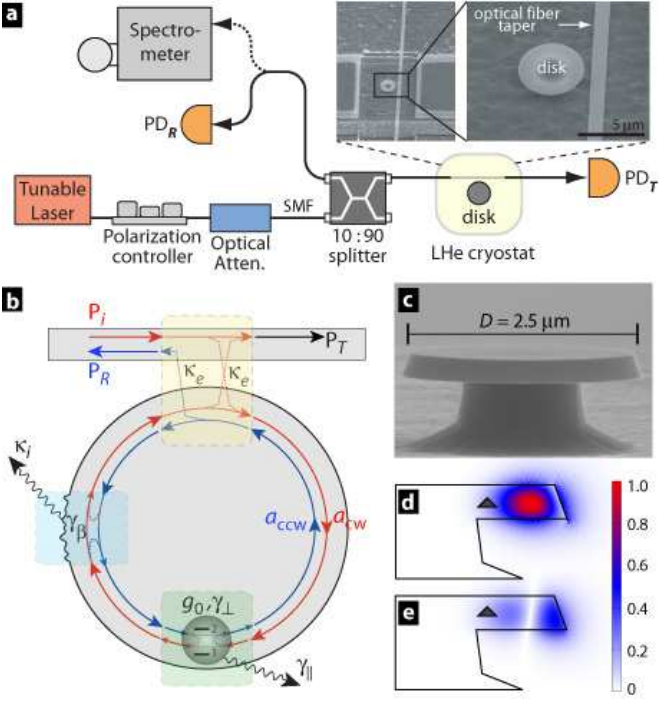}
\caption{\textbf{Schematics of the experimental apparatus and
system.} \textbf{a}, Diagram of the experimental setup showing a scanning electron microscope (SEM) image of a taper-coupled microdisk.  $\text{PD}_{R/T}$ are photodetectors for the
reflected/transmitted signals.  \textbf{b}, Illustration of the coupled microdisk-QD
system.  $a_{\text{CW/CCW}}$ are the amplitudes for the
clockwise/counterclockwise modes, $\text{P}_{i/R/T}$ are the
incident/reflected/transmitted signals, and $\kappa_{e}$ and $\kappa_{i}$ correspond to the fiber-to-cavity coupling and intrinsic cavity field decay rates, respectively. \textbf{c}, SEM image of one of the small microdisk cavities under study. \textbf{d}, FEM
simulations of the radial ($E_{\rho}$) and \textbf{e}, azimuthal ($E_{\phi}$)
electric field components of the TE$_{p=1,m=13}$ mode in
cross-section.  $p$ denotes the radial order and $m$ the azimuthal mode number.  The shaded triangle indicates the estimated QD position in this work.} \label{fig:schem_SEM_FEM}
\end{center}
\end{figure}

The experimental setup used is shown schematically in Fig.  \ref{fig:schem_SEM_FEM}(a)-(b).  At its core is a customized
liquid He cryostat\cite{ref:Srinivasan14} in which piezo-actuated stages have been integrated to incorporate optical
fiber taper testing while maintaining a sample temperature as low as $12$ K.  External cavity tunable lasers optically
pump the QD and probe the cavity-QD system near-resonance, and fused-fiber couplers direct the cavity's reflected and
transmitted signals to photodetectors and a spectrometer. The overall transmission of the fiber taper link is $50\%$ in
this work, and in many cases can be $\gtrsim90\%$, providing a very low-loss optical channel to probe the system. This
allows for the accurate estimation of quantities such as average intra-cavity photon number through measurement of the
resonant transmission of the taper waveguide when coupled to the cavity.

The system under investigation consists of InAs QDs embedded in a GaAs microdisk cavity.  The InAs QDs are grown in a
self-assembled manner with a density of $300-500$ ${\mu}m^{-2}$ on top of an InGaAs quantum well (a so-called
dot-in-a-well, or DWELL\cite{ref:Liu_G}).  The DWELL structure resides in the middle of a 256 nm thick GaAs layer that
forms the thin planar layer of the microdisk (see Fig. \ref{fig:schem_SEM_FEM}(c)).  Previous studies of this
material\cite{ref:Srinivasan15} indicate that isolated emission from single QDs at cryogenic temperature can be seen in
the wavelength range $\lambda=1290$-$1310$ nm, approximately $50$ nm red-shifted from the peak of the QD ensemble
emission.  Microdisks of diameter $D=2.5$ $\mu$m are created through electron beam lithography, plasma dry etching, and
wet undercut etching\cite{ref:Srinivasan9}. Finite-element-method (FEM) simulations (Fig.
\ref{fig:schem_SEM_FEM}(d-e)) of the microdisks show that the TE$_{1,13}$ whispering gallery mode (WGM) is resonant at
$\lambda\sim1300$ nm.  This optical mode has a radiation-limited quality factor $Q_{\text{rad}}>10^8$, and an effective
standing wave mode volume $V_{\text{sw}}=3.2(\lambda/n)^3$.  The peak coherent coupling rate for a QD excitonic state of
the type studied here (i.e., spontaneous emission lifetime $\tau_{\text{sp}}=1$ ns) with optimal placement and
dipole orientation is $g_{0}/2\pi=15$ GHz.  Since our QDs are not deterministically positioned in the cavity as in
recent studies\cite{ref:Badolato}, the actual exhibited coupling rate $g$ may be significantly smaller (see Methods).  The magnitude
of $g$ relative to the system decay rates, $\kappa_{T}$ (cavity field decay) and $\gamma_{\perp}$ (QD dephasing),
determines whether the system lies in the perturbative (weak coupling: $g<(\kappa_{T},\gamma_{\perp})$) or
non-perturbative (strong coupling: $g>(\kappa_{T},\gamma_{\perp})$) regime of cQED\cite{ref:Kimble2}.

\begin{figure}[t]
\begin{center}
\includegraphics[width=\linewidth]{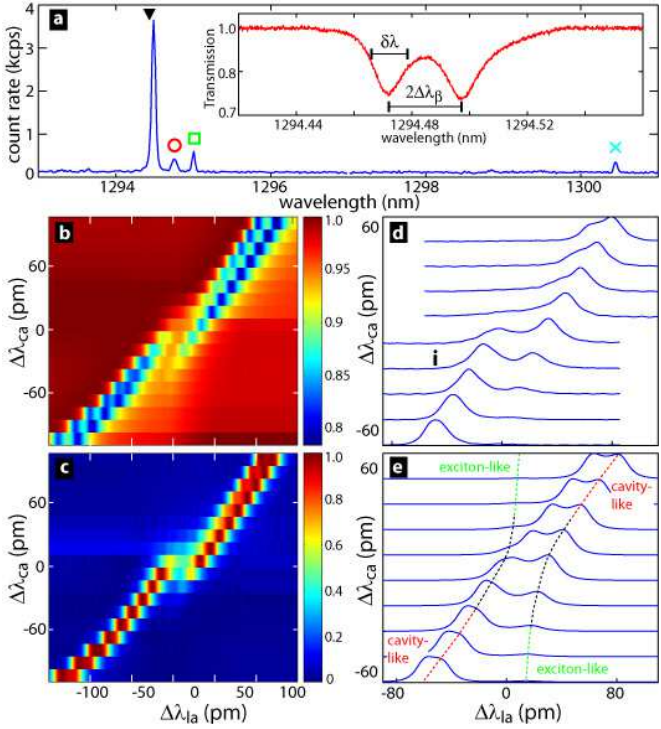}
\caption{\textbf{Reflection and transmission spectra from a
    strongly coupled microdisk-QD system}. \textbf{a}, Fiber-collected PL spectrum at a pump power of $30$ nW showing
  the cavity mode ($\blacktriangledown$), $X_{a}$ ($\textcolor{red}{\circ}$), $X_{b}$ ($\textcolor{green}{\square}$),
  and $X^{-}$ ($\textcolor{cyan}{\times}$) lines. The inset shows a transmission scan of the bare-cavity mode.  \textbf{b},
  Measured transmission and \textbf{c}, reflection spectra as a function of laser-QD detuning
  ($\Delta\lambda_{\text{la}}$) and cavity-QD detuning ($\Delta\lambda_{\text{ca}}$), where the cavity wavelength is
  tuned by the N$_{2}$ adsorption.  Transmission and reflection spectra are normalized to unity.  \textbf{d}, Experimental data and \textbf{e}, model plots for a series of reflected
  spectra in the central 120 pm region of cavity tuning.  The dashed lines in \textbf{e} are guides-to-the-eye for the exciton-like and cavity-like tuning.}
\label{fig:PL_plus_piezo_plus_cQED_schem}
\end{center}
\end{figure}

The process by which we identify a suitable device for studying cavity-QD coupling is described in the Methods section.
Figure \ref{fig:PL_plus_piezo_plus_cQED_schem}(a) shows the fiber-taper-collected PL spectrum from one such device that
has been cooled down to $15$ K.  Optical pumping of the QD is provided by exciting (also through the taper) a
blue-detuned higher-order WGM of the disk at $\lambda_{P}\sim982.2$ nm. The cavity mode, which is fed by background
emission processes\cite{ref:Hennessy3}, is the tall peak at the blue end of the spectrum. The three emission peaks red
of the cavity mode are the fine-structure-split\cite{ref:Forchel1} neutral single exciton lines, $X_{a}$ and $X_{b}$, and the negatively
charged single exciton line, $X^{-}$.



Further insight into the coupled cavity-QD system from PL are masked by the limited resolution of our spectrometer ($35$
pm). In this case the interesting behavior of the cavity-QD coupling can be studied by resonant spectroscopy of the
cavity mode using a fiber-coupled, narrowband (linewidth $<5$ MHz) tunable laser.  The inset to Fig.
\ref{fig:PL_plus_piezo_plus_cQED_schem}(a) shows the taper's transmission spectrum when it is placed in contact with the
side of the microdisk cavity and the cavity modes are detuned from the exciton lines.  As has been described in previous
work\cite{ref:Srinivasan9}, imperfections on the surface of the microdisk cause backscattering that couples the
initially degenerate traveling-wave WGMs.  If the backscattering rate $\gamma_{\beta}$ exceeds the total cavity loss
rate $\kappa_{T}$, this mode-coupling results in the formation of standing wave modes which are split in frequency.  The
transmission scan of Fig.  \ref{fig:PL_plus_piezo_plus_cQED_schem}(a) illustrates this effect in our system, with
TE$_{1,\pm13}$ modes appearing as a resonance doublet with splitting $2\Delta\lambda_{\beta}=31$ pm.  Each mode has a
linewidth of $\delta\lambda=13$ pm, corresponding to $Q=10^5$ and $\kappa_{T}/2\pi=1.2$ GHz.



To tune the cavity into resonance with the $X_{a}$ and $X_{b}$ exciton lines of the QD we introduce nitrogen (N$_{2}$)
gas into the cryostat\cite{ref:Mosor,ref:Srinivasan14}. As described in Ref. [\onlinecite{ref:Srinivasan14}] and in the
Methods section, this allows for continuous and repeated tuning over a 4 nm wavelength range of the cavity modes.  For
the first set of measurements, we operate with an input power of $470$ pW so that the system remains in a weak driving
limit with the estimated bare-cavity intracavity photon number $n_{\text{cav}}=0.03$.  The normalized transmission and
reflection spectra over a cavity tuning range of $240$ pm are displayed as a two-dimensional intensity image in Fig.
\ref{fig:PL_plus_piezo_plus_cQED_schem}(b)-(c).  Initially, we see a simple shift in the center wavelength of the cavity
doublet mode, but once the cavity mode frequency nears the transition frequency of the higher-energy exciton line
($X_{a}$) of the QD, the spectra change dramatically.  We see that coupling between the $X_{a}$-line and the cavity
modes results in a significant spectral splitting (vacuum Rabi splitting) that is evidenced in the characteristic
anti-crossing within both the transmitted and reflected signals. This anti-crossing is indicative of the cavity taking
on the character of the QD exciton, and vice versa, when the system becomes strongly coupled.  As the cavity is detuned
red of the $X_{a}$-line, the spectra regain their initial bare-cavity doublet shape.  Further tuning brings the cavity
modes into resonance with the $X_{b}$ exciton state.  Only a small frequency shift of the cavity modes (no
anti-crossing) is evident in this case, indicating that the $X_{b}$ state only weakly couples to either cavity mode.

Figure \ref{fig:PL_plus_piezo_plus_cQED_schem}(d) shows a series of reflection scans for a zoomed-in region of cavity
tuning, near where the $X_{a}$-line and the cavity are in resonance. In general, the character of these spectra are
complicated by the bimodal nature of WGM cavities. To adequately model the system, we use a quantum master equation
(QME) as presented in Ref.  [\onlinecite{ref:Srinivasan13}]. The model is used to solve for the steady-state reflected
and transmitted signals from the cavity as a function of parameters such as cavity-exciton coupling and excitonic
dephasing (the bare-cavity properties are known from detuned cavity spectra).  One other important parameter is the
relative phase, $\xi$, between the surface-scattering and exciton mode coupling.  The QD-cavity coupling strength with
the standing wave modes, $g_{\text{sw1,2}}$, is modified relative to that for traveling wave WGMs by a factor of
$(1{\pm}e^{i\xi})/\sqrt{2}$.

A series of reflected spectra produced by the model is shown in Fig.  \ref{fig:PL_plus_piezo_plus_cQED_schem}(e) for a
set of parameters, listed in Table \ref{tab:params}, which best estimates the measured reflected signal intensity,
exciton linewidth, relative coupling to the two standing wave modes, and anti-crossed splitting.  These parameters place
the $X_{a}$ exciton state and the TE$_{1,13}$ WGM in the good-cavity limit ($g>\gamma_{\perp}>\kappa_{T}$) of the strong
coupling regime.  We note that the achieved $g_{sw1}$ is about six times smaller than the maximum possible value based
on the cavity mode volume, and is likely due to the QD position being sub-optimal.  We estimate that the QD is located
$300$-$400$-nm inwards from the position of peak field strength of the TE$_{1,13}$ mode (see Fig.
\ref{fig:schem_SEM_FEM}(d)), with the dipole-moment of the $X_{a}$-line oriented radially and that of the $X_{b}$-line
oriented azimuthally.  This picture is consistent with the orthogonal $X_{a}$-$X_{b}$ polarization\cite{ref:Forchel1} and their relative measured coupling strengths.



\renewcommand{\arraystretch}{1.0}
\renewcommand{\extrarowheight}{1pt}
\begin{table*}
\caption{\textbf{Quantum master equation model parameters.}  See Methods for definition of $V_{\text{tw}}$ and $\eta$.}
\label{tab:params}
\begin{center}
\begin{tabular}{llllllllllll}
\hline
\hline
Parameter\hspace{20pt} & $V_{\text{tw}}$ &  $\eta$ & $\kappa_e/2\pi$ & $\kappa_i/2\pi$ & $\gamma_{\beta}/2\pi$ & $\xi$ & $\tau_{\text{rad}}$ & $g_{\text{sw1}}$ & $g_{\text{sw2}}$ & $\gamma_{\perp}/2\pi$ & $\gamma_{||}/2\pi$   \\
          & ($(\lambda/n)^3$) & & (GHz) & (GHz)  & (GHz) & (rad.) &  (ns) & (GHz) & (GHz) & (GHz) & (GHz) \\
\hline
Value & 6.4\hspace{25pt} & $0.21$\hspace{10pt} & $0.171$\hspace{10pt} & $0.91$\hspace{10pt}  & $1.99$\hspace{10pt} & $0.25\pi$\hspace{10pt} & $1$\hspace{15pt}  & $2.93$\hspace{10pt} & $1.21$\hspace{10pt} & $1.17$\hspace{15pt} & $0.55$  \\

\hline
\hline
\end{tabular}
\end{center}
\end{table*}
\renewcommand{\arraystretch}{1.0}
\renewcommand{\extrarowheight}{0pt}

The rate at which a single exciton can scatter incoming cavity photons is limited, resulting in a saturation in the
strongly-coupled QD-cavity response for large enough input power.  Two parameters used to characterize nonlinear
processes in cQED are the critical atom number $N_{0}$ and saturation photon number $m_{0}$, which gauge the number of
atoms needed to alter the cavity response and the number of photons needed to saturate the atomic transition,
respectively\cite{ref:Kimble2}.  These parameters are given by $N_{0}=2\kappa_{T}\gamma_{\perp}/g^2$ and
$m_{0}=\gamma_{\parallel}\gamma_{\perp}/4g^2$.  In our system $N_0=0.44$ and $m_0=0.02$ for the standing wave mode (sw1)
that couples most strongly to the QD.  This indicates that a single QD strongly affects the cavity response (which Fig.
\ref{fig:PL_plus_piezo_plus_cQED_schem} clearly indicates), while even an average intracavity photon number that is less
than one can saturate the QD response.

\begin{figure}[t]
\begin{center}
\includegraphics[width=\linewidth]{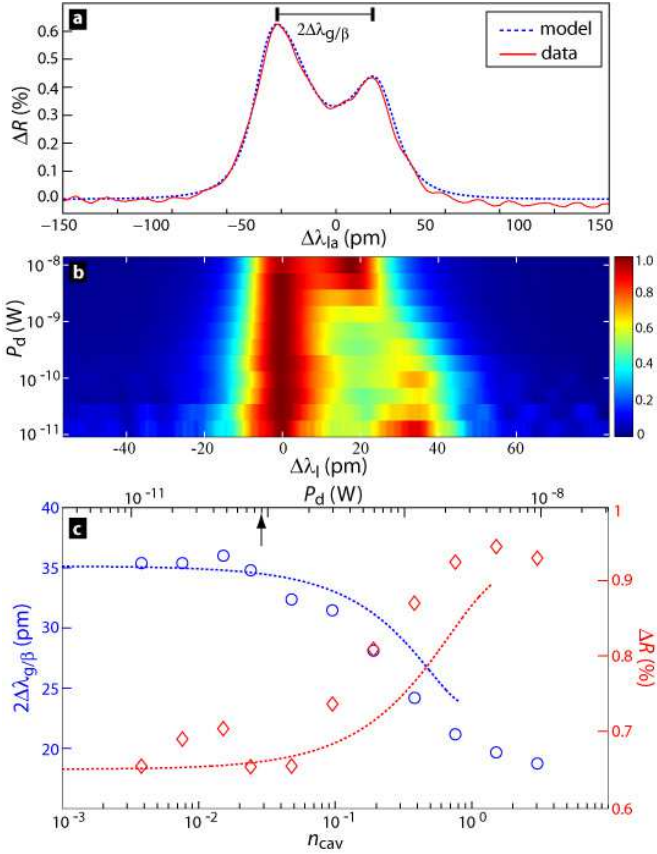}
\caption{\textbf{Power dependence of the QD-microcavity system}.
  \textbf{a}, Reflection spectrum from the QD-microdisk system near resonance (position (i) in Fig.
  \ref{fig:PL_plus_piezo_plus_cQED_schem}) under weak driving.  The solid red line is the measured reflected power normalized to input power; the dashed
  blue line is a QME model of the system. \textbf{b}, Normalized (to unity) reflected signal of panel a as a function of drive
  strength (dropped power in the bare-cavity, $P_{d}$) and detuning from the short-wavelength resonance peak
  ($\Delta\lambda_{l}$).  \textbf{c}, Measured and modeled saturation of the mode splitting and peak reflected signal
  level versus drive strength ($n_{\text{cav}}$ (bottom axis), $P_{d}$ (top axis)). The model is only plotted up to a
  drive power of $n_{\text{cav}}=1$ due to size limitations on the cavity mode Fock space which can be simulated.}
\label{fig:power_dependence}
\end{center}
\end{figure}

The measured power dependence of the QD-cavity system is shown in Fig.  \ref{fig:power_dependence}, where the cavity is
tuned into resonance with the $X_{a}$-line near the center of the anti-crossing region (scan marked '$\textbf{i}$' in
Fig.  \ref{fig:PL_plus_piezo_plus_cQED_schem}(d)), at which point the resonance peaks are nearly equal mixtures of
exciton and cavity mode.  Fig.  \ref{fig:power_dependence}(a) shows a plot of the measured reflected signal normalized
to input power ($\Delta R$) along with the modeled steady-state response of the cavity under weak driving conditions
($n_{\text{cav}}=0.03$).  As the input power to the cavity increases, Fig.  \ref{fig:power_dependence}(b) shows that the
spectral splitting due to cavity-QD interaction ($2\Delta\lambda_{g}$) begins to diminish as the exciton saturates, and
finally reaches a regime where the splitting is nearly two times smaller and due to surface-scattering
($2\Delta\lambda_{\beta}$).  Fig.  \ref{fig:power_dependence}(c) plots the resulting mode splitting
($2\Delta\lambda_{g/\beta}$) and peak $\Delta R$ as a function of the optical drive power.  Both the splitting and
reflected signal begin to saturate towards their bare-cavity values for $n_{\text{cav}}=0.1$.  The QME model (dotted
lines) predicts very similar behavior, albeit with a slightly higher drive power saturation point.  Both data and model,
however, show an extended saturation regime as expected due to the quantum fluctuations of a single
dipole\cite{ref:Savage}.  Such saturation behavior has previously been experimentally observed in atomic
systems\cite{ref:Hood2}.


Use of an optical-fiber-based waveguide to efficiently probe the microcavity-QD system opens up many interesting
possibilities for future devices and studies.  In particular, excitation and collection through the optical channel
allows for high resolution spectral and temporal studies of individual QD dynamics and a direct probe of the
intra-cavity field.  Studies of the quantum fluctuations of the strongly-coupled system\cite{ref:Birnbaum}, through
field and intensity correlations of the optical signal, are also now possible.  An immediate application is the creation
of an efficient fiber-coupled single photon source, while from a long-term perspective, it can be envisioned that the
fiber interface can serve as a means to transfer quantum information to and from the QD.  In comparison, atomic systems
have the considerable advantages of homogeneity, much lower dephasing, and an energy level structure compatible with
more complex manipulations of the quantum system.  Nitrogen-vacancy centers in diamond\cite{ref:Park_Y,ref:Santori3}
have been viewed as a system that can provide some of the beneficial aspects of cold atoms. The measurement apparatus
described here is equally applicable to this and other systems, and we are hopeful that it can be built upon to further
progress the development of solid-state cQED nodes with microchip-scalability.

\section*{Methods}

\begin{footnotesize}

  $\textbf{Device identification}$ A linear array of microdisk cavities is fabricated, with the disk diameter of $2.5
  \mu$m nominally held fixed.  Fluctuations in lithography cause the TE$_{1,13}$ mode wavelength to vary over a
  $1290$-$1310$ nm range.  We pump each device through the fiber taper and on resonance with one of its WGMs in the 980
  nm band\cite{ref:Srinivasan15}. This selectively excites QDs that lie in the disk periphery and overlap with the
  TE$_{1,13}$ mode.  For those devices in which isolated QD emission is observed, we examine the spectral position of
  the TE$_{1,13}$ mode relative to the QD states through PL and cavity transmission.  A digital wet etching
  process\cite{ref:Badolato} provides a cavity mode blue shift of 0.8 nm/cycle. This wet etch is repeated until the
  cavity mode lies blue (and within 1 nm) of the desired QD single exciton lines.  N$_{2}$ adsorption is then used to
  red-shift the mode into resonance with the desired exciton line.

  $\textbf{Transmission/Reflection measurements}$ The tunable laser used in transmission measurements provides a
  narrowband single mode laser line ($<5$ MHz) and continuous wavelength tuning through dithering of a piezo element.
  The transmitted and reflected signals are detected by TE-cooled (1 kHz bandwidth) and LN$_{2}$-cooled (150 Hz
  bandwidth) InGaAs  photodetectors, respectively.  In order to reduce detector noise the transmission and reflection signals
  are low-pass filtered (30 Hz cut-off) and the scans are averaged 10-20 times to produce the spectra of Fig.
  \ref{fig:PL_plus_piezo_plus_cQED_schem}.  PL is spectrally dispersed through a 550 mm Czerny-Turner spectrometer and
  detected on a 512 element LN-cooled InGaAs array (25 $\mu$m x 500 $\mu$m pixel size).  The effective spectrometer
  resolution is $35$ pm.

$\textbf{Cavity tuning}$ Nitrogen is released into the chamber
in discrete 5 second increments, with the flow rate adjusted so that a tuning level of $\sim$10 pm/step is achieved.
Once the shift is complete, the transmission and reflection spectra are acquired as described above.  At temperatures above $28$ K, the N$_2$ can
be removed from the disk surface and the cavity mode reset back to its original wavelength allowing for repeated tuning cycles.

$\textbf{Effective mode volume and g}$  The FEM-calculated traveling wave mode volume of the TE$_{1,13}$ WGM is $V_{\text{tw}}=6.4(\lambda/n)^3$.    The coherent coupling rate of the exciton to the traveling wave mode is $g_{\text{tw}} = \eta\sqrt{3c\lambda_{0}^2/8\pi{n^3}\tau_{\text{sp}}V_{\text{tw}}}$,  where $\eta$ accounts for the position and orientation of the exciton dipole ($\eta=1$ for an exciton dipole oriented parallel with, and positioned at, the peak of the cavity mode electric field).

  $\textbf{Quantum master equation simulations}$ Reference [\onlinecite{ref:Srinivasan13}] presents an appropriate model
  for our system.  We numerically solve the steady-state QME for the system's density matrix, from which
  the transmitted and reflected spectra from the cavity are generated.  A Fock space dimension of 6 for each cavity mode
  was used in modeling the drive power dependence of the system shown in Fig. \ref{fig:power_dependence}.  The
  expectation of the commutation between creation and annihilation operators for each mode was calculated to ensure accuracy of the
  simulation.

\end{footnotesize}

\section*{Acknowledgements}

\begin{footnotesize}

The authors thank Sanjay Krishna and Andreas Stintz of the Center
for High Technology Materials at the University of New Mexico for
providing material growth in support of this work.
\end{footnotesize}






\begin{thebibliography}{30}
\expandafter\ifx\csname natexlab\endcsname\relax\def\natexlab#1{#1}\fi
\expandafter\ifx\csname bibnamefont\endcsname\relax
  \def\bibnamefont#1{#1}\fi
\expandafter\ifx\csname bibfnamefont\endcsname\relax
  \def\bibfnamefont#1{#1}\fi
\expandafter\ifx\csname citenamefont\endcsname\relax
  \def\citenamefont#1{#1}\fi
\expandafter\ifx\csname url\endcsname\relax
  \def\url#1{\texttt{#1}}\fi
\expandafter\ifx\csname urlprefix\endcsname\relax\def\urlprefix{URL }\fi
\providecommand{\bibinfo}[2]{#2}
\providecommand{\eprint}[2][]{\url{#2}}

\bibitem[{\citenamefont{Kimble}(1998)}]{ref:Kimble2}
\bibinfo{author}{\bibfnamefont{H.~J.} \bibnamefont{Kimble}},
  \bibinfo{journal}{Physica Scripta} \textbf{\bibinfo{volume}{T76}},
  \bibinfo{pages}{127} (\bibinfo{year}{1998}).

\bibitem[{\citenamefont{Mabuchi and Doherty}(2002)}]{ref:Mabuchi}
\bibinfo{author}{\bibfnamefont{H.}~\bibnamefont{Mabuchi}} \bibnamefont{and}
  \bibinfo{author}{\bibfnamefont{A.~C.} \bibnamefont{Doherty}},
  \bibinfo{journal}{Science} \textbf{\bibinfo{volume}{298}},
  \bibinfo{pages}{1372} (\bibinfo{year}{2002}).

\bibitem[{\citenamefont{Hood et~al.}(1998)\citenamefont{Hood, Chapman, Lynn,
  and Kimble}}]{ref:Hood2}
\bibinfo{author}{\bibfnamefont{C.~J.} \bibnamefont{Hood}},
  \bibinfo{author}{\bibfnamefont{M.~S.} \bibnamefont{Chapman}},
  \bibinfo{author}{\bibfnamefont{T.~W.} \bibnamefont{Lynn}}, \bibnamefont{and}
  \bibinfo{author}{\bibfnamefont{H.~J.} \bibnamefont{Kimble}},
  \bibinfo{journal}{Phys. Rev. Lett.} \textbf{\bibinfo{volume}{80}},
  \bibinfo{pages}{4157} (\bibinfo{year}{1998}).

\bibitem[{\citenamefont{Hennrich et~al.}(2000)\citenamefont{Hennrich, Legero,
  Kuhn, and Rempe}}]{ref:Hennrich}
\bibinfo{author}{\bibfnamefont{M.}~\bibnamefont{Hennrich}},
  \bibinfo{author}{\bibfnamefont{T.}~\bibnamefont{Legero}},
  \bibinfo{author}{\bibfnamefont{A.}~\bibnamefont{Kuhn}}, \bibnamefont{and}
  \bibinfo{author}{\bibfnamefont{G.}~\bibnamefont{Rempe}},
  \bibinfo{journal}{Phys. Rev. Lett.} \textbf{\bibinfo{volume}{85}},
  \bibinfo{pages}{4872} (\bibinfo{year}{2000}).

\bibitem[{\citenamefont{Boca et~al.}(2004)\citenamefont{Boca, Miller, Birnbaum,
  Boozer, McKeever, and Kimble}}]{ref:Boca}
\bibinfo{author}{\bibfnamefont{A.}~\bibnamefont{Boca}},
  \bibinfo{author}{\bibfnamefont{R.}~\bibnamefont{Miller}},
  \bibinfo{author}{\bibfnamefont{K.~M.} \bibnamefont{Birnbaum}},
  \bibinfo{author}{\bibfnamefont{A.~D.} \bibnamefont{Boozer}},
  \bibinfo{author}{\bibfnamefont{J.}~\bibnamefont{McKeever}}, \bibnamefont{and}
  \bibinfo{author}{\bibfnamefont{H.~J.} \bibnamefont{Kimble}},
  \bibinfo{journal}{Phys. Rev. Lett.} \textbf{\bibinfo{volume}{93}},
  \bibinfo{pages}{233603} (\bibinfo{year}{2004}).

\bibitem[{\citenamefont{Keller et~al.}(2004)\citenamefont{Keller, Lange,
  Hayaska, Lange, and Walther}}]{ref:Keller}
\bibinfo{author}{\bibfnamefont{M.}~\bibnamefont{Keller}},
  \bibinfo{author}{\bibfnamefont{B.}~\bibnamefont{Lange}},
  \bibinfo{author}{\bibfnamefont{K.}~\bibnamefont{Hayaska}},
  \bibinfo{author}{\bibfnamefont{W.}~\bibnamefont{Lange}}, \bibnamefont{and}
  \bibinfo{author}{\bibfnamefont{H.}~\bibnamefont{Walther}},
  \bibinfo{journal}{Nature} \textbf{\bibinfo{volume}{431}},
  \bibinfo{pages}{1075} (\bibinfo{year}{2004}).

\bibitem[{\citenamefont{Aoki et~al.}(2006)\citenamefont{Aoki, Dayan, Wilcut,
  Bowen, Parkins, Kimble, Kippenberg, and Vahala}}]{ref:Aoki1}
\bibinfo{author}{\bibfnamefont{T.}~\bibnamefont{Aoki}},
  \bibinfo{author}{\bibfnamefont{B.}~\bibnamefont{Dayan}},
  \bibinfo{author}{\bibfnamefont{E.}~\bibnamefont{Wilcut}},
  \bibinfo{author}{\bibfnamefont{W.~P.} \bibnamefont{Bowen}},
  \bibinfo{author}{\bibfnamefont{A.~S.} \bibnamefont{Parkins}},
  \bibinfo{author}{\bibfnamefont{H.~J.} \bibnamefont{Kimble}},
  \bibinfo{author}{\bibfnamefont{T.~J.} \bibnamefont{Kippenberg}},
  \bibnamefont{and} \bibinfo{author}{\bibfnamefont{K.~J.}
  \bibnamefont{Vahala}}, \bibinfo{journal}{Nature}
  \textbf{\bibinfo{volume}{443}}, \bibinfo{pages}{671} (\bibinfo{year}{2006}).

\bibitem[{\citenamefont{Colombe et~al.}(2007)\citenamefont{Colombe, Steinmetz,
  Dubois, Linke, Hunger, and Reichel}}]{ref:reichel1}
\bibinfo{author}{\bibfnamefont{Y.}~\bibnamefont{Colombe}},
  \bibinfo{author}{\bibfnamefont{T.}~\bibnamefont{Steinmetz}},
  \bibinfo{author}{\bibfnamefont{G.}~\bibnamefont{Dubois}},
  \bibinfo{author}{\bibfnamefont{F.}~\bibnamefont{Linke}},
  \bibinfo{author}{\bibfnamefont{D.}~\bibnamefont{Hunger}}, \bibnamefont{and}
  \bibinfo{author}{\bibfnamefont{J.}~\bibnamefont{Reichel}}
  (\bibinfo{year}{2007}), \bibinfo{note}{quant-ph/0706.1390}.

\bibitem[{\citenamefont{Park et~al.}(2006)\citenamefont{Park, Cook, and
  Wang}}]{ref:Park_Y}
\bibinfo{author}{\bibfnamefont{Y.-S.} \bibnamefont{Park}},
  \bibinfo{author}{\bibfnamefont{A.~K.} \bibnamefont{Cook}}, \bibnamefont{and}
  \bibinfo{author}{\bibfnamefont{H.}~\bibnamefont{Wang}},
  \bibinfo{journal}{Nano Letters} \textbf{\bibinfo{volume}{6}},
  \bibinfo{pages}{2075} (\bibinfo{year}{2006}).

\bibitem[{\citenamefont{Reithmaier et~al.}(2004)\citenamefont{Reithmaier, Sek,
  Loffer, Hoffman, Kuhn, Reitzenstein, Keldysh, Kulakovskii, Reinecke, and
  Forchel}}]{ref:Reithmaier}
\bibinfo{author}{\bibfnamefont{J.~P.} \bibnamefont{Reithmaier}},
  \bibinfo{author}{\bibfnamefont{G.}~\bibnamefont{Sek}},
  \bibinfo{author}{\bibfnamefont{A.}~\bibnamefont{Loffer}},
  \bibinfo{author}{\bibfnamefont{C.}~\bibnamefont{Hoffman}},
  \bibinfo{author}{\bibfnamefont{S.}~\bibnamefont{Kuhn}},
  \bibinfo{author}{\bibfnamefont{S.}~\bibnamefont{Reitzenstein}},
  \bibinfo{author}{\bibfnamefont{L.~V.} \bibnamefont{Keldysh}},
  \bibinfo{author}{\bibfnamefont{V.~D.} \bibnamefont{Kulakovskii}},
  \bibinfo{author}{\bibfnamefont{T.~L.} \bibnamefont{Reinecke}},
  \bibnamefont{and} \bibinfo{author}{\bibfnamefont{A.}~\bibnamefont{Forchel}},
  \bibinfo{journal}{Nature} \textbf{\bibinfo{volume}{432}},
  \bibinfo{pages}{197} (\bibinfo{year}{2004}).

\bibitem[{\citenamefont{Yoshie et~al.}(2004)\citenamefont{Yoshie, Scherer,
  Hendrickson, Khitrova, Gibbs, Rupper, Ell, Schenkin, and
  Deppe}}]{ref:Yoshie3}
\bibinfo{author}{\bibfnamefont{T.}~\bibnamefont{Yoshie}},
  \bibinfo{author}{\bibfnamefont{A.}~\bibnamefont{Scherer}},
  \bibinfo{author}{\bibfnamefont{J.}~\bibnamefont{Hendrickson}},
  \bibinfo{author}{\bibfnamefont{G.}~\bibnamefont{Khitrova}},
  \bibinfo{author}{\bibfnamefont{H.}~\bibnamefont{Gibbs}},
  \bibinfo{author}{\bibfnamefont{G.}~\bibnamefont{Rupper}},
  \bibinfo{author}{\bibfnamefont{C.}~\bibnamefont{Ell}},
  \bibinfo{author}{\bibfnamefont{Q.}~\bibnamefont{Schenkin}}, \bibnamefont{and}
  \bibinfo{author}{\bibfnamefont{D.}~\bibnamefont{Deppe}},
  \bibinfo{journal}{Nature} \textbf{\bibinfo{volume}{432}},
  \bibinfo{pages}{200} (\bibinfo{year}{2004}).

\bibitem[{\citenamefont{Peter et~al.}(2005)\citenamefont{Peter, Senellart,
  Martrou, Lema$\hat{\text{i}}$tre, Hours, G\'{e}rard, and Bloch}}]{ref:Peter}
\bibinfo{author}{\bibfnamefont{E.}~\bibnamefont{Peter}},
  \bibinfo{author}{\bibfnamefont{P.}~\bibnamefont{Senellart}},
  \bibinfo{author}{\bibfnamefont{D.}~\bibnamefont{Martrou}},
  \bibinfo{author}{\bibfnamefont{A.}~\bibnamefont{Lema$\hat{\text{i}}$tre}},
  \bibinfo{author}{\bibfnamefont{J.}~\bibnamefont{Hours}},
  \bibinfo{author}{\bibfnamefont{J.~M.} \bibnamefont{G\'{e}rard}},
  \bibnamefont{and} \bibinfo{author}{\bibfnamefont{J.}~\bibnamefont{Bloch}},
  \bibinfo{journal}{Phys. Rev. Lett.} \textbf{\bibinfo{volume}{95}},
  \bibinfo{pages}{067401} (\bibinfo{year}{2005}).

\bibitem[{\citenamefont{Hennessy et~al.}(2007)\citenamefont{Hennessy, Badolato,
  Winger, Gerace, Atature, Guide, Falt, Hu, and Imamoglu}}]{ref:Hennessy3}
\bibinfo{author}{\bibfnamefont{K.}~\bibnamefont{Hennessy}},
  \bibinfo{author}{\bibfnamefont{A.}~\bibnamefont{Badolato}},
  \bibinfo{author}{\bibfnamefont{M.}~\bibnamefont{Winger}},
  \bibinfo{author}{\bibfnamefont{D.}~\bibnamefont{Gerace}},
  \bibinfo{author}{\bibfnamefont{M.}~\bibnamefont{Atature}},
  \bibinfo{author}{\bibfnamefont{S.}~\bibnamefont{Guide}},
  \bibinfo{author}{\bibfnamefont{S.}~\bibnamefont{Falt}},
  \bibinfo{author}{\bibfnamefont{E.}~\bibnamefont{Hu}}, \bibnamefont{and}
  \bibinfo{author}{\bibfnamefont{A.}~\bibnamefont{Imamoglu}},
  \bibinfo{journal}{Nature (London)} \textbf{\bibinfo{volume}{445}},
  \bibinfo{pages}{896} (\bibinfo{year}{2007}).

\bibitem[{\citenamefont{Press et~al.}(2007)\citenamefont{Press, Gotzinger,
  Reitzenstein, Hoffmann, Loffler, Kamp, Forchel, and Yamamoto}}]{ref:Press}
\bibinfo{author}{\bibfnamefont{D.}~\bibnamefont{Press}},
  \bibinfo{author}{\bibfnamefont{S.}~\bibnamefont{Gotzinger}},
  \bibinfo{author}{\bibfnamefont{S.}~\bibnamefont{Reitzenstein}},
  \bibinfo{author}{\bibfnamefont{C.}~\bibnamefont{Hoffmann}},
  \bibinfo{author}{\bibfnamefont{A.}~\bibnamefont{Loffler}},
  \bibinfo{author}{\bibfnamefont{M.}~\bibnamefont{Kamp}},
  \bibinfo{author}{\bibfnamefont{A.}~\bibnamefont{Forchel}}, \bibnamefont{and}
  \bibinfo{author}{\bibfnamefont{Y.}~\bibnamefont{Yamamoto}},
  \bibinfo{journal}{Phys. Rev. Lett.} \textbf{\bibinfo{volume}{98}},
  \bibinfo{pages}{117402} (\bibinfo{year}{2007}).

\bibitem[{\citenamefont{Michler et~al.}(2000)\citenamefont{Michler, Kiraz,
  Becher, Schoenfeld, Petroff, Zhang, Hu, and Imamoglu}}]{ref:Michler}
\bibinfo{author}{\bibfnamefont{P.}~\bibnamefont{Michler}},
  \bibinfo{author}{\bibfnamefont{A.}~\bibnamefont{Kiraz}},
  \bibinfo{author}{\bibfnamefont{C.}~\bibnamefont{Becher}},
  \bibinfo{author}{\bibfnamefont{W.~V.} \bibnamefont{Schoenfeld}},
  \bibinfo{author}{\bibfnamefont{P.~M.} \bibnamefont{Petroff}},
  \bibinfo{author}{\bibfnamefont{L.}~\bibnamefont{Zhang}},
  \bibinfo{author}{\bibfnamefont{E.}~\bibnamefont{Hu}}, \bibnamefont{and}
  \bibinfo{author}{\bibfnamefont{A.}~\bibnamefont{Imamoglu}},
  \bibinfo{journal}{Science} \textbf{\bibinfo{volume}{290}},
  \bibinfo{pages}{2282} (\bibinfo{year}{2000}).

\bibitem[{\citenamefont{Santori et~al.}(2002)\citenamefont{Santori, Fattal,
  Vuckovic, Solomon, and Yamamoto}}]{ref:Santori2}
\bibinfo{author}{\bibfnamefont{C.}~\bibnamefont{Santori}},
  \bibinfo{author}{\bibfnamefont{D.}~\bibnamefont{Fattal}},
  \bibinfo{author}{\bibfnamefont{J.}~\bibnamefont{Vuckovic}},
  \bibinfo{author}{\bibfnamefont{G.}~\bibnamefont{Solomon}}, \bibnamefont{and}
  \bibinfo{author}{\bibfnamefont{Y.}~\bibnamefont{Yamamoto}},
  \bibinfo{journal}{Nature} \textbf{\bibinfo{volume}{419}},
  \bibinfo{pages}{594} (\bibinfo{year}{2002}).

\bibitem[{\citenamefont{Birnbaum et~al.}(2005)\citenamefont{Birnbaum, Boca,
  Miller, Boozer, Northup, and Kimble}}]{ref:Birnbaum}
\bibinfo{author}{\bibfnamefont{K.~M.} \bibnamefont{Birnbaum}},
  \bibinfo{author}{\bibfnamefont{A.}~\bibnamefont{Boca}},
  \bibinfo{author}{\bibfnamefont{R.}~\bibnamefont{Miller}},
  \bibinfo{author}{\bibfnamefont{A.}~\bibnamefont{Boozer}},
  \bibinfo{author}{\bibfnamefont{T.~E.} \bibnamefont{Northup}},
  \bibnamefont{and} \bibinfo{author}{\bibfnamefont{H.~J.}
  \bibnamefont{Kimble}}, \bibinfo{journal}{Nature}
  \textbf{\bibinfo{volume}{436}}, \bibinfo{pages}{87} (\bibinfo{year}{2005}).

\bibitem[{\citenamefont{Knight et~al.}(1997)\citenamefont{Knight, Cheung,
  Jacques, and Birks}}]{ref:Knight}
\bibinfo{author}{\bibfnamefont{J.~C.} \bibnamefont{Knight}},
  \bibinfo{author}{\bibfnamefont{G.}~\bibnamefont{Cheung}},
  \bibinfo{author}{\bibfnamefont{F.}~\bibnamefont{Jacques}}, \bibnamefont{and}
  \bibinfo{author}{\bibfnamefont{T.~A.} \bibnamefont{Birks}},
  \bibinfo{journal}{Opt. Lett.} \textbf{\bibinfo{volume}{22}},
  \bibinfo{pages}{1129} (\bibinfo{year}{1997}).

\bibitem[{\citenamefont{Spillane et~al.}(2003)\citenamefont{Spillane,
  Kippenberg, Painter, and Vahala}}]{ref:Spillane2}
\bibinfo{author}{\bibfnamefont{S.~M.} \bibnamefont{Spillane}},
  \bibinfo{author}{\bibfnamefont{T.~J.} \bibnamefont{Kippenberg}},
  \bibinfo{author}{\bibfnamefont{O.~J.} \bibnamefont{Painter}},
  \bibnamefont{and} \bibinfo{author}{\bibfnamefont{K.~J.}
  \bibnamefont{Vahala}}, \bibinfo{journal}{Phys. Rev. Lett.}
  \textbf{\bibinfo{volume}{91}}, \bibinfo{pages}{043902}
  (\bibinfo{year}{2003}).

\bibitem[{\citenamefont{Srinivasan et~al.}(2004)\citenamefont{Srinivasan,
  Barclay, Borselli, and Painter}}]{ref:Srinivasan7}
\bibinfo{author}{\bibfnamefont{K.}~\bibnamefont{Srinivasan}},
  \bibinfo{author}{\bibfnamefont{P.~E.} \bibnamefont{Barclay}},
  \bibinfo{author}{\bibfnamefont{M.}~\bibnamefont{Borselli}}, \bibnamefont{and}
  \bibinfo{author}{\bibfnamefont{O.}~\bibnamefont{Painter}},
  \bibinfo{journal}{Phys. Rev. B} \textbf{\bibinfo{volume}{70}},
  \bibinfo{pages}{081306R} (\bibinfo{year}{2004}).

\bibitem[{\citenamefont{Srinivasan et~al.}(2005)\citenamefont{Srinivasan,
  Borselli, Johnson, Barclay, Painter, Stintz, and Krishna}}]{ref:Srinivasan9}
\bibinfo{author}{\bibfnamefont{K.}~\bibnamefont{Srinivasan}},
  \bibinfo{author}{\bibfnamefont{M.}~\bibnamefont{Borselli}},
  \bibinfo{author}{\bibfnamefont{T.~J.} \bibnamefont{Johnson}},
  \bibinfo{author}{\bibfnamefont{P.~E.} \bibnamefont{Barclay}},
  \bibinfo{author}{\bibfnamefont{O.}~\bibnamefont{Painter}},
  \bibinfo{author}{\bibfnamefont{A.}~\bibnamefont{Stintz}}, \bibnamefont{and}
  \bibinfo{author}{\bibfnamefont{S.}~\bibnamefont{Krishna}},
  \bibinfo{journal}{Appl. Phys. Lett.} \textbf{\bibinfo{volume}{86}},
  \bibinfo{pages}{151106} (\bibinfo{year}{2005}).

\bibitem[{\citenamefont{Srinivasan and
  Painter}(2007{\natexlab{a}})}]{ref:Srinivasan14}
\bibinfo{author}{\bibfnamefont{K.}~\bibnamefont{Srinivasan}} \bibnamefont{and}
  \bibinfo{author}{\bibfnamefont{O.}~\bibnamefont{Painter}},
  \bibinfo{journal}{Appl. Phys. Lett.} \textbf{\bibinfo{volume}{90}},
  \bibinfo{pages}{031114} (\bibinfo{year}{2007}{\natexlab{a}}).

\bibitem[{\citenamefont{Liu et~al.}(2000)\citenamefont{Liu, Stintz, Li, Newell,
  Gray, Varangis, Malloy, and Lester}}]{ref:Liu_G}
\bibinfo{author}{\bibfnamefont{G.~T.} \bibnamefont{Liu}},
  \bibinfo{author}{\bibfnamefont{A.}~\bibnamefont{Stintz}},
  \bibinfo{author}{\bibfnamefont{H.}~\bibnamefont{Li}},
  \bibinfo{author}{\bibfnamefont{T.~C.} \bibnamefont{Newell}},
  \bibinfo{author}{\bibfnamefont{A.~L.} \bibnamefont{Gray}},
  \bibinfo{author}{\bibfnamefont{P.~M.} \bibnamefont{Varangis}},
  \bibinfo{author}{\bibfnamefont{K.~J.} \bibnamefont{Malloy}},
  \bibnamefont{and} \bibinfo{author}{\bibfnamefont{L.~F.}
  \bibnamefont{Lester}}, \bibinfo{journal}{IEEE J. Quan. Elec.}
  \textbf{\bibinfo{volume}{36}}, \bibinfo{pages}{1272} (\bibinfo{year}{2000}).

\bibitem[{\citenamefont{Srinivasan et~al.}(2007)\citenamefont{Srinivasan,
  Painter, Stintz, and Krishna}}]{ref:Srinivasan15}
\bibinfo{author}{\bibfnamefont{K.}~\bibnamefont{Srinivasan}},
  \bibinfo{author}{\bibfnamefont{O.}~\bibnamefont{Painter}},
  \bibinfo{author}{\bibfnamefont{A.}~\bibnamefont{Stintz}}, \bibnamefont{and}
  \bibinfo{author}{\bibfnamefont{S.}~\bibnamefont{Krishna}}
  (\bibinfo{year}{2007}), \bibinfo{note}{physics/0706.1831}.

\bibitem[{\citenamefont{Badolato et~al.}(2005)\citenamefont{Badolato, Hennessy,
  Atature, Dreiser, Hu, Petroff, and Imamoglu}}]{ref:Badolato}
\bibinfo{author}{\bibfnamefont{A.}~\bibnamefont{Badolato}},
  \bibinfo{author}{\bibfnamefont{K.}~\bibnamefont{Hennessy}},
  \bibinfo{author}{\bibfnamefont{M.}~\bibnamefont{Atature}},
  \bibinfo{author}{\bibfnamefont{J.}~\bibnamefont{Dreiser}},
  \bibinfo{author}{\bibfnamefont{E.}~\bibnamefont{Hu}},
  \bibinfo{author}{\bibfnamefont{P.~M.} \bibnamefont{Petroff}},
  \bibnamefont{and} \bibinfo{author}{\bibfnamefont{A.}~\bibnamefont{Imamoglu}},
  \bibinfo{journal}{Science} \textbf{\bibinfo{volume}{308}},
  \bibinfo{pages}{1158} (\bibinfo{year}{2005}).

\bibitem[{\citenamefont{Kulakovski et~al.}(1999)\citenamefont{Kulakovski,
  Bacher, , Weigand, Kummell, Forchel, Borovitskaya, Leonardi, and
  Hommel}}]{ref:Forchel1}
\bibinfo{author}{\bibfnamefont{V.~D.} \bibnamefont{Kulakovski}},
  \bibinfo{author}{\bibfnamefont{G.}~\bibnamefont{Bacher}}, ,
  \bibinfo{author}{\bibfnamefont{R.}~\bibnamefont{Weigand}},
  \bibinfo{author}{\bibfnamefont{T.}~\bibnamefont{Kummell}},
  \bibinfo{author}{\bibfnamefont{A.}~\bibnamefont{Forchel}},
  \bibinfo{author}{\bibfnamefont{E.}~\bibnamefont{Borovitskaya}},
  \bibinfo{author}{\bibfnamefont{K.}~\bibnamefont{Leonardi}}, \bibnamefont{and}
  \bibinfo{author}{\bibfnamefont{D.}~\bibnamefont{Hommel}},
  \bibinfo{journal}{Phys. Rev. Lett.} \textbf{\bibinfo{volume}{82}},
  \bibinfo{pages}{1780} (\bibinfo{year}{1999}).

\bibitem[{\citenamefont{Mosor et~al.}(2005)\citenamefont{Mosor, Hendrickson,
  Richards, Sweet, Khitrova, Gibbs, Yoshie, Scherer, Shchekin, and
  Deppe}}]{ref:Mosor}
\bibinfo{author}{\bibfnamefont{S.}~\bibnamefont{Mosor}},
  \bibinfo{author}{\bibfnamefont{J.}~\bibnamefont{Hendrickson}},
  \bibinfo{author}{\bibfnamefont{B.~C.} \bibnamefont{Richards}},
  \bibinfo{author}{\bibfnamefont{J.}~\bibnamefont{Sweet}},
  \bibinfo{author}{\bibfnamefont{G.}~\bibnamefont{Khitrova}},
  \bibinfo{author}{\bibfnamefont{H.}~\bibnamefont{Gibbs}},
  \bibinfo{author}{\bibfnamefont{T.}~\bibnamefont{Yoshie}},
  \bibinfo{author}{\bibfnamefont{A.}~\bibnamefont{Scherer}},
  \bibinfo{author}{\bibfnamefont{O.~B.} \bibnamefont{Shchekin}},
  \bibnamefont{and} \bibinfo{author}{\bibfnamefont{D.~G.} \bibnamefont{Deppe}},
  \bibinfo{journal}{Appl. Phys. Lett.} \textbf{\bibinfo{volume}{87}},
  \bibinfo{pages}{141105} (\bibinfo{year}{2005}).

\bibitem[{\citenamefont{Srinivasan and
  Painter}(2007{\natexlab{b}})}]{ref:Srinivasan13}
\bibinfo{author}{\bibfnamefont{K.}~\bibnamefont{Srinivasan}} \bibnamefont{and}
  \bibinfo{author}{\bibfnamefont{O.}~\bibnamefont{Painter}},
  \bibinfo{journal}{Phys. Rev. A} \textbf{\bibinfo{volume}{75}},
  \bibinfo{pages}{023814} (\bibinfo{year}{2007}{\natexlab{b}}).

\bibitem[{\citenamefont{Savage and Carmichael}(1988)}]{ref:Savage}
\bibinfo{author}{\bibfnamefont{C.~M.} \bibnamefont{Savage}} \bibnamefont{and}
  \bibinfo{author}{\bibfnamefont{H.~J.} \bibnamefont{Carmichael}},
  \bibinfo{journal}{IEEE J. Quan. Elec.} \textbf{\bibinfo{volume}{24}},
  \bibinfo{pages}{1495} (\bibinfo{year}{1988}).

\bibitem[{\citenamefont{Santori et~al.}(2006)\citenamefont{Santori, Tamarat,
  Neumann, Wrachtrup, Fattal, Beausoleil, Rabeau, Olivero, Greentree, Prawer
  et~al.}}]{ref:Santori3}
\bibinfo{author}{\bibfnamefont{C.}~\bibnamefont{Santori}},
  \bibinfo{author}{\bibfnamefont{P.}~\bibnamefont{Tamarat}},
  \bibinfo{author}{\bibfnamefont{P.}~\bibnamefont{Neumann}},
  \bibinfo{author}{\bibfnamefont{J.}~\bibnamefont{Wrachtrup}},
  \bibinfo{author}{\bibfnamefont{D.}~\bibnamefont{Fattal}},
  \bibinfo{author}{\bibfnamefont{R.}~\bibnamefont{Beausoleil}},
  \bibinfo{author}{\bibfnamefont{J.}~\bibnamefont{Rabeau}},
  \bibinfo{author}{\bibfnamefont{P.}~\bibnamefont{Olivero}},
  \bibinfo{author}{\bibfnamefont{A.}~\bibnamefont{Greentree}},
  \bibinfo{author}{\bibfnamefont{S.}~\bibnamefont{Prawer}},
  \bibnamefont{et~al.}, \bibinfo{journal}{Phys. Rev. Lett.}
  \textbf{\bibinfo{volume}{97}}, \bibinfo{pages}{247401}
  (\bibinfo{year}{2006}).

\end{thebibliography}

\end{document}